\documentclass{article}

\usepackage{arxiv}

\usepackage{amsmath}		   

\usepackage[utf8]{inputenc} 
\usepackage[T1]{fontenc}    
\usepackage{hyperref}       
\usepackage{url}            
\usepackage{booktabs}       
\usepackage{amsfonts}       
\usepackage{nicefrac}       
\usepackage{microtype}      
\usepackage{cleveref}       
\usepackage{lipsum}         
\usepackage{graphicx}
\usepackage{natbib}
\usepackage{doi}

\usepackage{xcolor}

\title{Flock2: A model for orientation-based social flocking}

\date{April 26, 2024}

\newif\ifuniqueAffiliation

\uniqueAffiliationtrue

\ifuniqueAffiliation 

\author{ \href{https://orcid.org/0000-0002-0449-981X}{\includegraphics[scale=0.06]{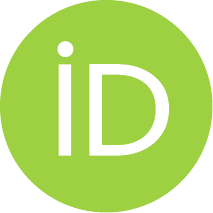}\hspace{1mm}Rama Carl Hoetzlein} \\
	Quanta Sciences\\
	Ithaca, NY 14850 \\
	\texttt{ramahoetzlein@gmail.com} \\
}
	
\else

\fi

\renewcommand{\shorttitle}{\textit{Flock2} A model for orientation-based social flocking}

\hypersetup{
pdftitle={Flock2: A model for orientation-based social flocking},
pdfsubject={q-bio.QM, q-bio.NC},
pdfauthor={Rama Carl Hoetzlein},
pdfkeywords={flocking, agitation waves, collective escape, bird flight, aerodynamics, boids, modeling, computational biology},
}

\begin{document}
\maketitle

\begin{abstract}
The aerial flocking of birds, or murmurations, has fascinated observers while presenting many challenges to behavioral study and simulation. We examine how the periphery of murmurations remain well bounded and cohesive. We also investigate agitation waves, which occur when a flock is disturbed, developing a plausible model for how they might emerge spontaneously. To understand these behaviors a new model is presented for orientation-based social flocking. Previous methods model inter-bird dynamics by considering the neighborhood around each bird, and introducing forces for avoidance, alignment, and cohesion as three dimensional vectors that alter acceleration. Our method introduces orientation-based social flocking that treats social influences from neighbors more realistically as a desire to turn, indirectly controlling the heading in an aerodynamic model. While our model can be applied to any flocking social bird we simulate flocks of starlings, \textit{Sturnus vulgaris}, and demonstrate the possibility of orientation waves in the absence of predators. Our model exhibits spherical and ovoidal flock shapes matching observation. Comparisons of our model to Reynolds' on energy consumption and frequency analysis demonstrates more realistic motions, significantly less energy use in turning, and a plausible mechanism for emergent orientation waves.
\end{abstract}


\section{Introduction}
Collective group formation is well known as many animals aggregate into dense groups for protection. Fish form schools as a defense \citep{Nikolsky1955} \citep{Breder1959} \citep{Pitcher1983} \citep{Godin1985} \citep{Magurran1985} \citep{Parrish2002} and grazing mammals such as oxen and sheep exhibit strong inclinations to aggregate at all times for protection \citep{Hamilton1971} \citep{Treves2000}. The concept of the selfish herd, introduced by Hamilton, provides one theoretical basis: many species aggregate out of self interest as a form of protection \citep{Hamilton1971}. This attractive force to seek cover is balanced by the competing need to maintain some distance between individuals \citep{Emlen1952}.

Flocking is a form of collective animal behavior in birds that is strongly influenced by the predator-prey relationship. Empirical studies have shown that flocks are larger, flocking for longer times, and with more activity and vigilance in monitoring when predators are present \citep{Caraco1980} \citep{Carbone2003} \citep{Procaccini2011} \citep{Goodenough2017}. Birds such as quelea, juncos and starlings will defend from external predators through aggregation into murmurations \citep{Vine1971} \citep{Caraco1980}. Other studies have noted that time budgets, the way a bird divides its time between feeding, monitoring and flocking, are correlated with predators \citep{McNamara1992} \citep{Sirot2006}.

Aerial flocking is generally believed to be an airborne extension of predator defense in combination with activities for foraging and roosting \citep{Siegfried1975} \citep{Feare1984} \citep{Cresswell1994} \citep{Beauchamp2004} \citep{Fischl1987} \citep{Tinbergen1981} \citep{Caccamise1991}. Specific aerial patterns such as escape waves, whereby oscillations are induced in a flock at the point of attack, have been observed in relation to predators \citep{Procaccini2011}. The geometric shape of an aerial flock is influenced by many factors including season, temperature, food resources and predator presence \citep{Carere2009}.

Computational modeling has become an important tool in the understanding of aerial flocking. The work of \citet{Couzin2002} on group formation for fish and birds in three dimensions is based in part on an earlier model by \citet{Reynolds1987}. Reynolds defines avoidance as the tendency for a bird to avoid collisions with neighbors, alignment as the tendency for a bird to match direction and velocity with neighbors, and cohesion as the tendency to move toward the local average position of neighbors. This approach, relying on vector forces, continues to be the primary method employed in subsequent models of aerial flocking dynamics \citep{Couzin2002} \citep{Couzin2005} \citep{Parrish2005} \citep{Buhl2006} \citep{Hemelrijk2008} \citep{Hildenbrandt2010}.

Despite the extensive use of Reynolds' model, there are discrepancies between observational and computational biology that our work seeks to address. Some models introduce a preferred direction \citep{Couzin2005} or roosting site \citep{Hildenbrandt2010} to account for the fact that additional terms are needed in Reynolds' model to maintain cohesively bounded flocks. These are generally unsatisfying as a continuous in-flight adjustment since roosting is a resting behavior \citep{Caccamise1990}.

\begin{figure*}[!htb]
  \includegraphics[width=\textwidth]{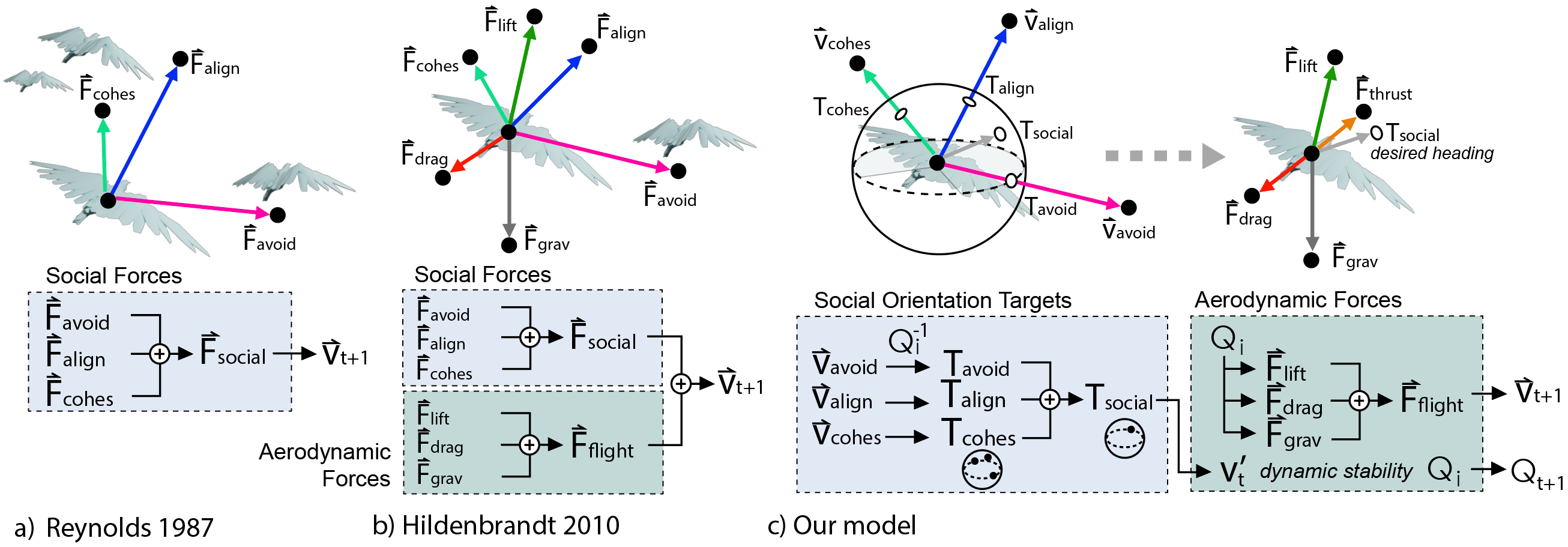}  
  \caption{Previous methods implement social factors as 3D vector forces in a) \citet{Reynolds1987} and b) with aerodynamic forces in \citet{Hildenbrandt2010}. We develop a flocking model c) in which bird motion is strictly controlled by an aerodynamic model for which the social behaviors of avoidance, alignment and cohesion are combined to act as a single actuator for a desired change in orientation.}
  \label{fig_models}
\end{figure*}

\citet{Hildenbrandt2010} model aerial flocking by combining the social forces of Reynolds' model with aerodynamic forces of flight, even though the only group of physical forces considered in the study of real bird flight are aerodynamic since the Reynolds' social forces are not actual forces of mechanical work \citep{Pennycuick1968} \citep{Withers1981}. The social influences are perceptual, which presumably influence the bird to change its wing shape, direction and speed.

Agitation waves, whereby a predator causes the flock to deform or escape, have been observed extensively in starlings \citep{Carere2009} \citep{Procaccini2011} \citep{Storms2023}. \citet{Hemelrijk2015} develop a computational model for agitation waves, yet this required the explicit coding of a "transmission mechanism" for escape messages from one bird to the next, and whereby a predator is strictly assumed to trigger the first message. 

These discrepancies between observational and computational biology led us to consider a new foundational model in which social forces are observational rather than explicit 3D vectors doing mechanical work. This approach is more realistic since a bird may sense its neighbors but is ultimately only able to alter its wing shape, heading and speed.

\clearpage

\section{Model and simulation}
\label{model}

This work is motivated by the following observations:

\begin{itemize}
\item Real birds are known to turn to a greater degree than altering their speed for energy savings. "Speed is a stiffer mode than orientation, as it is more costly for a bird to change its speed than its heading.” \citep{Cavagna2010}

\item Bird physiology provides that sensory information, including the vision of neighbors, is integrated to control flight. \citep{Altshuler2015}

\item Social forces in Reynold's model are not actual forces doing mechanical work, they are better thought of as perceptual factors that influence heading.

\item Computational models for the aerodynamics of birds must allow for some articulation or heading control, since this is the only way a bird can alter its desired direction.

\end{itemize}

These observations led to the consideration of an orientation-based social model which controls an actuated aerodynamic model. An important aspect of this design is that only the total aerodynamic forces are permitted to update the bird's velocity and position, whereas the social orientation-model is only permitted to influence the desired heading. 

\subsection{Bird Flight}
\label{bird_flight}

Understanding the flight of real birds should ideally provide the basis for flocking models. As we introduce an aerodynamic model coupled to a flocking model, the study of real bird flight is essential in model design and parameter analysis. Real bird flight has been analyzed either as a flapping wing \citep{Pennycuick1968} \citep{Rayner1993} \citep{Rayner1999} or as a fixed wing \citep{Withers1981} \citep{Norberg1985}.

A distinction is made between mechanical power for the total external forces of motion \citep{Pennycuick1968} \citep{Dial1997}, and metabolic power which is the total bodily energy produced including basal metabolism, respiration, internal resistance and transformations of chemical energy \citep{Daan1990} \citep{Rayner2001} \citep{Ward2001} \citep{Ward2004} \citep{Tobalske2007}. 

We consider only mechanical power in our approach whereas a more complete model would simulate both the internal and external functions of a bird. The theory of flapping wings suggests that mechanical power output follow a U-shaped curve as real birds use the dynamic stall effect to achieve better lift at low speeds and with drag forces dominating at higher speeds \citep{Pennycuick1968}. Our analysis examines simulated birds in light of this theory.

The starling is used as reference for our models with an average mass of 80 grams and typical mechanical power output of 10 W flying at an average speed of 10 m/s \citep{Withers1981} \citep{Dial1997} \citep{Nudds2000} \citep{Schmidt2008}. The predominant energy use corresponds to the lift force applied downward to move a volume of air with a reactionary upward force on the body. The next most energetic expenditure is to overcome drag including profile drag on the wings, with turbulence, and parasitic drag on the body. Studies show that energy usage increases proportionally with bird weight \citep{Schmidt2008}. The remaining mechanical energy is applied to accelerate, decelerate and turn.  We rely on this knowledge for both model design and analysis.

\subsection{Mathematical background}
\label{math_background}

In Reynolds' model the social terms for avoidance, alignment and cohesion are summed to give the total force acting on the motion of the bird:
\begin{equation}
F_{social} = F_{avoidance} + F_{alignment} + F_{cohesion}
\end{equation}	

These social forces act directly on velocity as if they were doing mechanical work with velocity updating position \citep{Reynolds1987}. 

Aerodynamics have also been incorporated into social flocking in birds by \citet{Hildenbrandt2010}. In their approach, the Reynolds' social forces are combined in the same way using vectors to give $F_{social}$. With a separate term, they model the bird as a fixed wing where lift is the force resulting from differential pressure flowing over the wing, and drag is the fluid resistance of the air against forward motion. These are summed with gravity to give $F_{flight}$, the total aerodynamic forces as a 3D vector. The social forces are scaled by a constant factor $k$ so their magnitude is similar to the flight forces, and added together to give the final bird motion:
\begin{equation}
F_{total} = F_{flight} + k F_{social} 
\end{equation}	

Their work was the first to model a distinction between social forces, which are based on bird behavior in response to neighbors or predators, and aerodynamic forces, which are based on the in-flight forces of lift, drag, thrust and gravity.  Fig. \ref{fig_models}b provides a summary of this approach.

Many recent models for both fish and birds rely on Reynold's vector-based model \citep{Couzin2002} \citep{Erra2004} \citep{Hemelrijk2008} \citep{Hemelrijk2011} \citep{Hemelrijk2012} \citep{Hemelrijk2015} \citep{Silva2009} \citep{Hildenbrandt2010} \citep{Costanzo2021} \citep{Papadopoulou2023}. Only a few models of flocking take other approaches, such as fuzzy logic \citep{Bajec2005} \citep{Demsar2014}, phase-transition physics \citep{Vicsek2012} or shape morphing \citep{Ho2012}. 

\begin{figure}
  \includegraphics[width=0.5\textwidth]{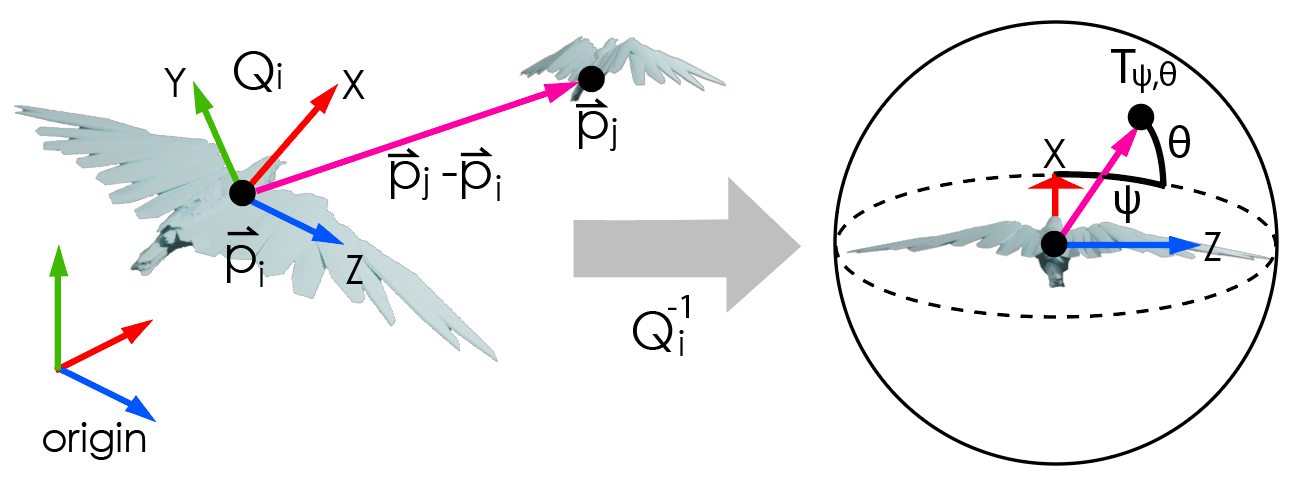}  
  \caption{Given bird at position $p_i$ with orientation expressed as a quaternion $Q_i$, we use the inverse quaternion $Q^{-1}_i$ to transform the direction of a neighboring bird $p_j$ from world space into the local coordinate frame, allows us to find the yaw $\psi_i$ and pitch $\theta_i$ angles for the direction of a given target to attract or repel.}
  \label{fig_orient}
\end{figure}

\subsection{Model design}
\label{model_design}

We introduce a new simulation model, Flock2 (or \textit{Floids}), for social flocking with flight. Our model consists of a high-level orientation model for social factors which controls a low-level aerodynamic model. The aerodynamic model, Flightsim, is fully integrated into Flock2 and provides actuated control over the orientation of a fixed wing-body with lift, drag and gravity. Although this is not a flapping model it is an improvement over previous methods for aerodynamics in social flocking. An orientation model is implemented that more realistically models the ability of a bird to turn rather than accelerate in a specific direction. 

\subsection{Modeling social factors as orientation goals}
\label{social_factors}

The social factors of avoidance, alignment and cohesion in our system are implemented as orientation targets, with the orientation of each bird maintained as a quaternion $Q_i$. Given a bird $i$ at position $p_i$ the avoidance term considers the direction to the nearest neighbor j. The directional vector $p_j - p_i$ is transformed into the local reference frame of the bird using the inverse quaternion $Q^{-1}_i$, as in Figure \ref{fig_orient}.

Avoidance
\begin{equation}  
    \overrightarrow{v}_{avoid} = (\mathbf{p}_j - \mathbf{p}_i)
\end{equation}
\begin{equation}
    T_{\psi_i}^{avoid} = k_{avoid} \cdot \tan^{-1}\left( \overrightarrow{v}_{avoid} \: Q^{-1}_i \right) \frac{1}{d},
\end{equation}
\begin{equation}
    T_{\theta_i}^{avoid} = k_{avoid} \cdot \sin^{-1}\left( \overrightarrow{v}_{avoid} \: Q^{-1}_i \right) \frac{1}{d},
\end{equation}

Within the local reference frame, the arctangent and arcsine give the local yaw $\psi_i$ and pitch $\theta_i$ angles of the neighboring bird j with respect to the forward direction of bird i. This can be understood as the direction to bird j projected onto the sphere of vision of the current bird (see Fig. \ref{fig_orient}). When searching for neighbors we only consider those within a limited forward field-of-view. 

Alignment considers the direction indicated by the average local velocity $v_j$ for the $n_i$ neighbors of bird i.
  
\begin{equation}  
    \overrightarrow{v}_{align} = \frac{\sum_j\mathbf{v}_j}{n_i}
\end{equation}
\begin{equation}
    T_{\psi_i}^{align} = k_{align} \cdot \tan^{-1} \left( \overrightarrow{v}_{align} \: Q^{-1}_i \right),
\end{equation}
\begin{equation}
    T_{\theta_i}^{align} = k_{align} \cdot \sin^{-1} \left( \overrightarrow{v}_{align} \: Q^{-1}_i \right),
\end{equation}

For cohesion we find the direction given by the average local position, or centroid, of the $n_i$ neighbors of bird i. 

\begin{equation}
    \overrightarrow{v}_{cohesion} = \left(\frac{\sum_j\mathbf{p}_j}{n_i} - \mathbf{p}_i \right)
\end{equation}
\begin{equation}
    T_{\psi_i}^{cohesion} = k_{coh} \cdot \tan^{-1} \left( \overrightarrow{v}_{cohesion} \: Q^{-1}_i \right),
\end{equation}
\begin{equation}
    T_{\theta_i}^{cohesion} = k_{coh} \cdot \sin^{-1}  \left( \overrightarrow{v}_{cohesion} \: Q^{-1}_i \right),
\end{equation}

The constants $k_{align}$ and $k_{coh}$ express positive attraction for alignment and cohesion respectively, whereas the parameter $k_{avoid}$ is negative to express the strength of avoidance.

In each of these sets of equations the 3D vectors for the nearest bird $\overrightarrow{v}_{avoid}$, the average neighborhood velocity $\overrightarrow{v}_{align}$, and the neighborhood centroid $\overrightarrow{v}_{cohesion}$, are identical to those used in the Reynolds' force model. The key difference is that, rather than using these as forces directly, we then project these vectors onto the visual sphere of the bird to derive orientation targets $T_{\phi_i}, T_{\theta_i}$ for each term (see Fig. \ref{fig_orient})

\begin{figure}
  \includegraphics[width=\textwidth]{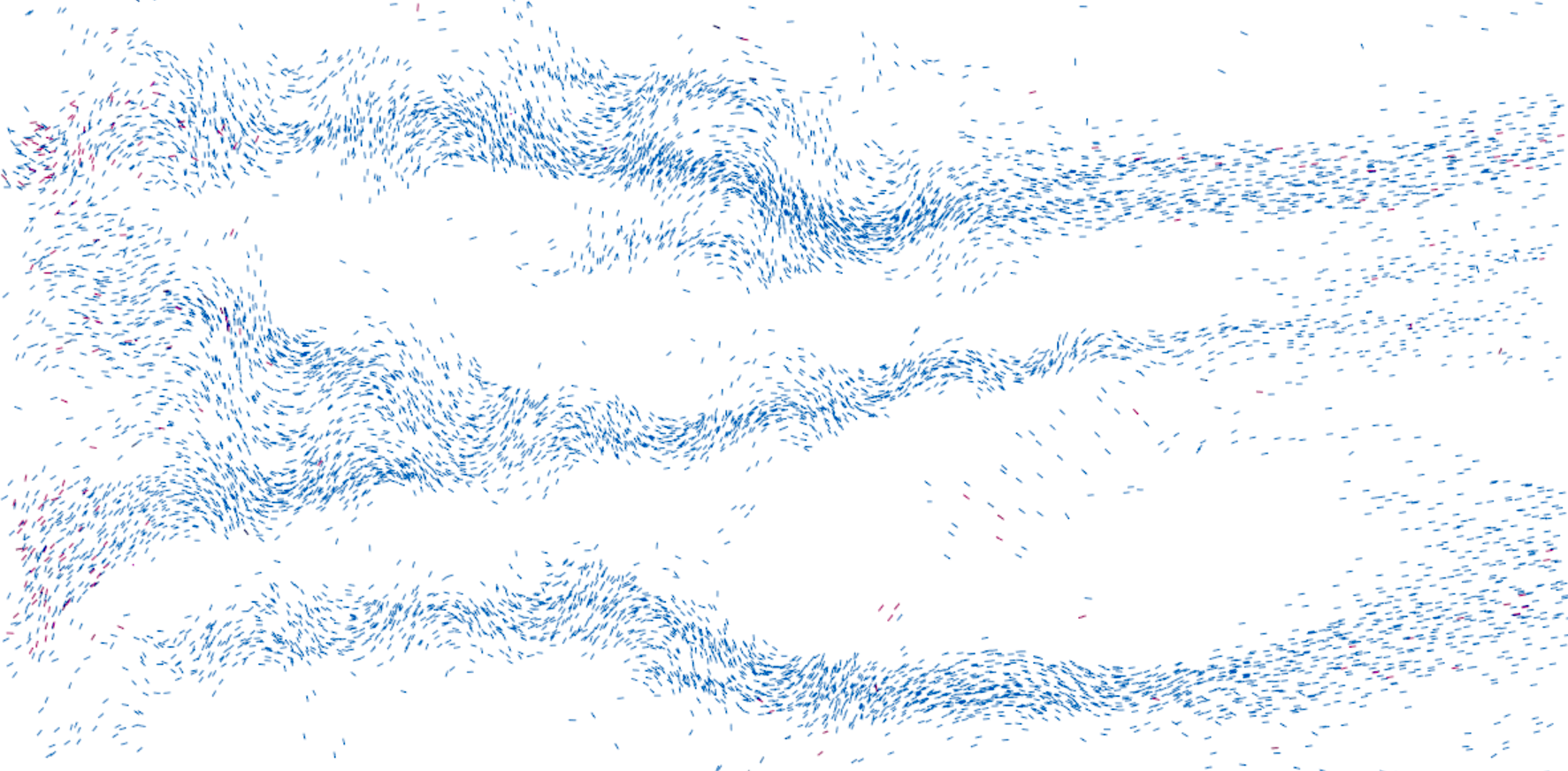}  
  \caption{With only the three terms for avoidance, alignment and cohesion, and without roosting or peripheral bounds, simulations produce snake-like formations rather than spherical or ovoid patterns. We introduce a peripheral bound for birds on the margin to seek cover within the flock. This applies to any bird which is more isolated and is distinct from the simulation domain bounds. See section \ref{boundary_term}. }
  \label{fig_unbound}
\end{figure}

\subsection{Peripheral boundary term}
\label{boundary_term}

Boundary cohesion is also an important feature of murmurations that we investigated further. In starlings the flock boundaries are highly structured and coherent with few outlying birds. In previous models an additional term for attraction to a preferred direction \citep{Couzin2005} or to a roosting site \citep{Hildenbrandt2010} was often introduced to aid in grouping. This is generally unsatisfying as a continuous in-flight model since roosting is a ground behavior. \citet{Caccamise1990} found that most birds are attracted more to feeding areas called diurnal activity centers (DACs). Birds nest for the night at a communal roost, which may be near a DAC, yet European starlings are “less faithful to roost sites than to their DACs.” They may commute 3 to 12 km to 12 different roosts \citep{Caccamise1990}. Neither feeding DACs nor roosts are satisfactory explanations for the strong, near instantaneous in-flight boundary cohesion of murmurations. Yet some peripheral cohesion is required to present more spherical or ovoidal shaped flocks.

In models containing only the three classical Reynolds' terms of avoidance, alignment and cohesion, we found these terms alone do not form densely packed globular or ovoidal flocks with many birds (Fig. \ref{fig_unbound}), which are the commonly observed in real flocks \citep{Carere2009}.

Marginal predation provides a behavioral basis for modeling peripheral birds \citep{Hirsch2011} \citep{Duffield2017}. The theory of marginal predation suggests that birds on the periphery are more exposed in-flight, and therefore have a greater evolved tendency and inclination to return to the flock. \citet{Hildenbrandt2010} explore peripheral coherence by measuring the degree of centrality $C_i$ within the local topological group, where the direction they use for this adjustment is still the Reynolds’ social force for cohesion.

We model peripheral birds as having a rapid, in-flight, evolved response based on maintaining self-cover, whereby they not only cohere together more strongly, but avoid the periphery by plunging into the overall flock. This is modeled as a distinct social orientation term. Those in the center of the flock are well protected all around so they only need avoid neighbors and stay aligned. On the periphery, birds may sense that they are becoming isolated and turn back toward the center. This evolved response is plausible based on an evolutionary response to marginal predation, and does not require the explicit presence of a predator.

The degree to which a bird is peripheral in the flock can be estimated from the neighborhood density 0 < $n_i/B$ < 1, where B is an adjustable constant defining the thickness of the border region. For these birds, we turn them back toward the flock centroid $C$ with a strength based on density.  

\begin{equation}
    \overrightarrow{v}_{bound} = C - \mathbf{p_i}
\end{equation}
\begin{equation}
    T_{\psi_i}^{bound} = k_{bound} \left( \frac{B - n_i}{B} \right) \cdot \tan^{-1} \left( \overrightarrow{v}_{bound} \: Q^{-1}_i \right),
\end{equation}
\begin{equation}
    T_{\theta_i}^{bound} = k_{bound} \left( \frac{B - n_i}{B} \right) \cdot \sin^{-1} \left( \overrightarrow{v}_{bound} \: Q^{-1}_i \right),
\end{equation}

When $n_i = 0$ the bird has no neighbors in the field of vision and the peripheral turning strength is 1. When $n_i = B$ the bird has maximal neighbors and the peripheral turning strength becomes 0.

\subsection{Total orientation target}
\label{total_orientation}

The input to the aerodynamic model is a target orientation T that expresses the direction that any bird wishes to fly in yaw and pitch, as shown in Figure \ref{fig_models}c. To determine the total desired target orientation of bird i, we sum the turning influence from all social factors.

\begin{equation}
     T_{\psi_i,\theta_i}^{total} = T_{\psi,\theta}^{avoid} + T_{\psi,\theta}^{align} + T_{\psi,\theta}^{cohesion} + T_{\psi,\theta}^{bound}
\end{equation}
    
Care must be taken since the sum takes place on a sphere. Yaw $\psi$ values must be permitted to wrap around +/- 180 degrees from forward, while pitch $\theta$ values wrap around +/- 90 degrees.
    
\subsection{Aerodynamic model}
\label{aerodynamic}

Real bird flight is analyzed with fixed wing models, or more commonly an aerodynamic flapping model \citep{Norberg1985} \citep{Pennycuick1968}. With respect to simulating flocks of birds, the model by \citet{Hildenbrandt2010} was the first to introduce aerodynamic flight based on fixed wings. In their model the body and wings are always oriented along the forward velocity $v_i$, and the social forces $F_{social}$ are not coupled to the aerodynamic forces $F_{flight}$. Nonetheless, their model was the first to exhibit lift loss when banking, whole flock turning, and vertical speed changes (climbing and diving).

Typically the bird is oriented along the direction of forward velocity $v_i$ \citep{Couzin2002} \citep{Hildenbrandt2010} \citep{Hemelrijk2011} \citep{Reynolds1987}. On the other hand, a complete aerodynamic model for fixed wing flight should include angular velocity, spin, rotational inertia, and torque generated by each lift surface \citep{Yechout2014}. Our flocking model relies on Flightsim, a fixed wing model that resides between these, providing a level of control in which a target orientation deflects the current heading, and the dynamic stability of the wing reorients the moving body along the direction of motion with a delay. 

At each time step the current orientation $Q_i$ defines the local axes for the forward (F), up (U) and side (S) directions.

\begin{equation}
  F = <1,0,0>Q_i, U = <0,1,0>Q_i, S = <0,0,1>Q_i
\end{equation}

Deflection of the forward direction F' is computed from the input orientation target $T_{\psi_i, \theta_i}$ by constructing quaternions that represent the turning rate around the up (U) and side (S) axes respectively:
 
\begin{equation}
  Q_{\psi} = Q_{angleaxis} \left( (dt/k_r) T_{\psi_i}^{total}, U \right)
\end{equation}
\begin{equation}
  Q_{\theta} = Q_{angleaxis} \left( (dt/k_r) T_{\theta_i}^{total}, S \right)
\end{equation}
\begin{equation}
  F' = F Q_{\psi} Q_{\theta}
\end{equation}
The operator $Q_{angleaxis}$ constructs a quaternion given an angle and axis. F' represents the modified forward direction vector toward the desired target, whose angle with F depends on the reaction rate. At least two factors influence the flight response rate of birds: (1) the reaction time of the vision system which is measured to be around 76 ms \citep{Pomeroy1977}, and (2) the rate at which a starling can turn to a desired orientation. Since a measured value of the latter is unknown we simplify the turning reaction speed $k_r$ as a single parameter: time to achieve a desired orientation, with 250 ms used in our simulations.
  
Lift and drag are defined by the aerodynamic equations:

\begin{equation}
  F_{lift} = \frac{1}{2} \rho v^2 A C_L
\end{equation}
\begin{equation}
  F_{drag} = \frac{1}{2} \rho v^2 A C_D
\end{equation}

where $\rho$ is the air density, A is the wing area, $v$ is the airflow velocity (bird velocity along F plus wind), and $C_L$ is the coefficient of lift. We assume a constant weight, wing area and air density similar to \citet{Hildenbrandt2010}, with one notable exception: the angular difference between the modified forward direction F' and the body velocity $v$ of the wing allows us to define the wing angle-of-attack in the classical sense. 

\begin{equation}
  aoa = cos^{-1} ( F' \cdot v )
\end{equation}

The angle of attack is used to lookup a variable coefficient of lift $C_L$ defined by a standard wing profile with stalls occurring at higher angles if power is not increased. Due to wing shape and flapping, birds are better able to avoid stalls so this factor can be adjusted or eliminated \citep{Dhawan1991}.

Thrust is applied in the direction of motion. Realistically this would be caused by flapping but is treated here as a constant force in the direction of motion.

\begin{equation}
  F_{thrust} = \hat{v} \cdot k_{power}
\end{equation}

Forces for lift, drag, gravity and thrust are combined as 3D vectors to give the total aerodynamic forces acting on the body.

\begin{equation}
   F_{flight} = F_{lift} + F_{drag} + F_{thrust} + F_{gravity}
\end{equation}

The principle of dynamic stability states that a deflected aerodynamic body will naturally want to re-orient along its direction of motion, not instantaneously, but with greater strength as the deflection increases. The final step in the aerodynamic model applies this principle to fractionally re-orient the bird $Q_i$ from the targeted direction F' toward the body velocity $v_i$ with the weighting factor $k_s$. In the following equation the $Q_{fromto}$ operator constructs a quaternion rotating one vector to another with fraction k.

\begin{equation}
  Q_i^{t+1} = Q_i^t \cdot Q_{fromto} \left( F', v_i, k_s \right)
\end{equation}

While the base aerodynamic model was intended for fixed wing aircraft we assume that birds are capable of much greater control. In our model birds are assumed to be able to adjust their roll, pitch and yaw rapidly at will merely by changing wing shape.

\subsubsection{Time integration}
\label{integration}

The complete system including both the social factors and aerodynamic model is shown in Figure \ref{fig_models}c. These are evaluated for each bird and integrated at each time step. The neighbor search is performed first using efficient grid-based acceleration techniques for every bird to find its seven topological neighbors. For all birds, the current bird positions are used to find the desired target orientations $T_{\psi_i, \theta_i}$ based on the social factors - this constitutes the bird's perceptual activity. These orientation targets update the aerodynamic model to give the total aerodynamic force $F_{flight}$ and the new orientation $Q_i^{t+1}$ - this constitutes the bird's mechanical activity. Lastly, position and velocity are integrated with Euler's method to update the bird for the next time step.
  
\begin{equation}
   v_i^{t+1} = v_i^t + \left( F_{flight} / mass \right) dt
\end{equation}  
\begin{equation}
   p_i^{t+1} = p_i^t + v_i^t dt
\end{equation}  

Note that only the $F_{flight}$ forces from the aerodynamic model are permitted to update the bird velocities and positions. 

\section{Results}
\label{results}

Our model produced emergent behaviors exhibiting many characteristics of real flocks that were not explicitly programmed, including non-penetrating flock collisions, vortex formations, plausible splitting and merging, and both low and high frequency turn propagation. We define spontaneous orientation waves as visible, low frequency waves that traverse a flock whether they occur with a predator or not. Orientation waves were especially surprising to observe in our model since they were initially not expected, not explicitly programmed (emergent), and occurred in the absence of any predefined trigger such as a predator.

Consistent with real flocks our introduction of the boundary term allowed for the spherical and ovoidal pattern to be observed most frequently (Fig. \ref{fig_patterns}a). However, formations of miniflocks, singletons, and indefinite shapes could be also observed within individual simulation runs (Fig. \ref{fig_patterns}). This is elaborated in the discussion section.

Multiple experiments were conducted to better understand our model. To quantify our model in relation to others we performed an aggregate energy analysis of bird flight, section \ref{energy}. Since our primary contribution is an orientation-based model, we perform a frequency and spectrum power analysis to better understand the oscillations produced during orientation waves, section \ref{frequency}. We also perform sensitivity analysis with multiple simulations over isolated model parameters, section \ref{sensitivity}. Finally, we conducted experiments between two flocks to test the response rate of simulated birds to head-on collisions, section \ref{collisions}.

Observational research of real three-dimensional flocking data by \citet{Ballerini2008a} found that starlings typically respond to between six or seven topological neighbors. In all experiments for both Flock2 and Reynolds' model, we use a finite, topological search with seven neighbors for the three primary social terms. With the boundary term, this is extended to a finite visual support radius of 10 m in order to determine peripheral bird density, requiring $O(kN)$ complexity. The simulation bounds are 200 $m^2$, with wrap-around conditions horizontally and avoidance for ground and ceiling.

We study the Flock2 model with the boundary term (F2) in comparison to Reynolds' model (REY). Since existing models are all derived from Reynolds' classical vector-based force model, we choose to compare against this original work. The model of \citet{Hildenbrandt2010} includes aerodynamic forces, but in their work these are separated from the social forces we wish to examine with respect to energy. We simulated between 5,000 and 20,000 birds in our starling model over a 200 $m^2$ region, with the number of birds and simulation runs described in each experiment type.

\begin{figure}[ht]
  \begin{center}
  \includegraphics[width=0.6\textwidth]{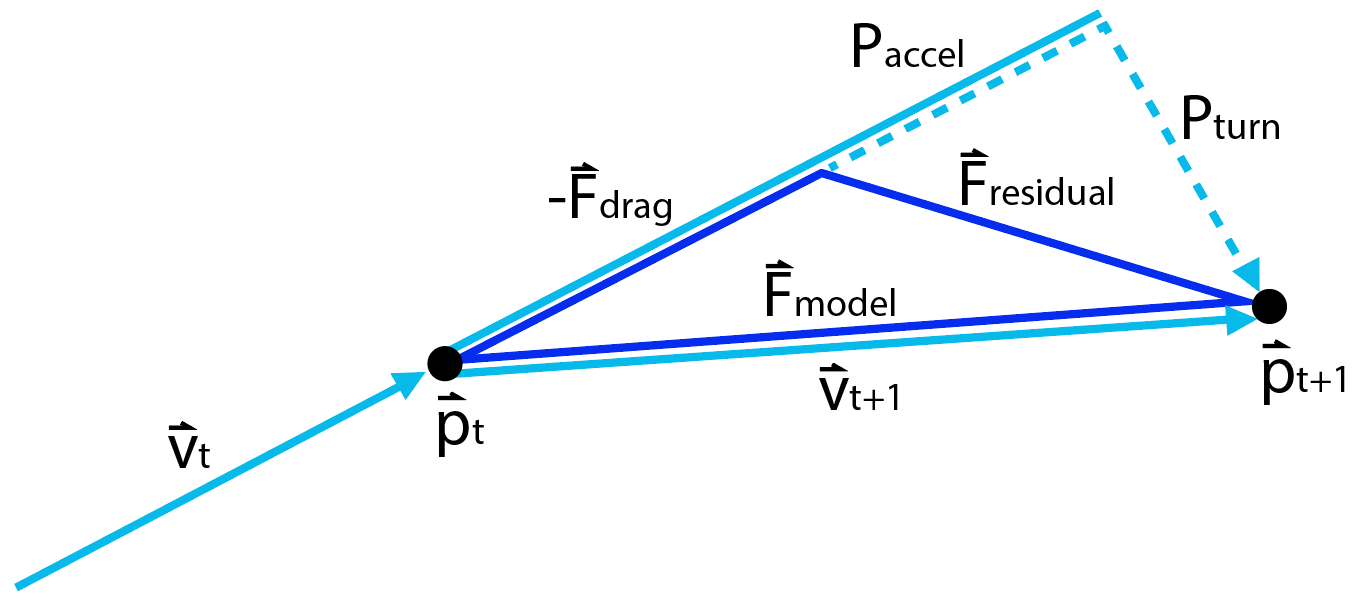}  
  \end{center}
  \caption{Residual forces are found to analyze simulated energies in any model of flight. A certain portion of energy is applied to overcome drag, $-F_{drag}$. The residual $F_{residual}$ is the remaining amount of force which is applied to bird motion after lift, drag and gravity have been subtracted. Of this, the forward component of work along the previous velocity direction $v_t$ is applied to accelerate or decelerate the bird, $P_{accel}$. The side component, $P_{turn}$, is applied to turn the bird. In this figure we assume dt=1 and mass=1 for clarity.}
  \label{fig_residual}
\end{figure}

\begin{table}
  \begin{center}
  \label{tab:results}
  \begin{tabular}{lrrrr}
  \toprule
  Output & F2 (ours) & REY & REY : F2 & \multicolumn{1}{c}{REY}  \\
  & \multicolumn{1}{c}{watts} & watts & \multicolumn{1}{c}{\%} & \% of $P_{mech}$ \\ 
\midrule
$P_{lift}$ & 7.80 & 8.30 & 106\% & 76.54\% \\
$P_{drag}$ & 2.62 & 2.51 & 95\% & 23.15\% \\
$P_{accel}$ & 0.0035 & 0.0020 & 56\% & 0.02\% \\
$P_{turn}$ & 0.00008 & 0.0320 & 39500\% & 0.30\% \\
$P_{mech}$ & 10.42 & 10.84 & 104\% & 100.00\% \\
Ave. Speed & 10.02 & 10.18 & 101\% &  \\
  \bottomrule
  \end{tabular}
  \end{center}
  \vspace{1em}
  \caption{Energy comparison of bird output in mechanical work averaged over 10,000 birds for the Flock2 and Reynolds models. Simulations are calibrated using equilibrium parameters before analysis. }
\end{table}

\subsection{Energy analysis}
\label{energy}
Simulations were conducted to compare energy analysis between our model (F2) and Reynolds' model (REY). As our model was designed in standard international units (SI) we are able to compute mechanical energy in watts. To calibrate both models we consider the equilibrium condition of steady horizontal flight. 

Mechanical power $P_{mech}$ is the energy required to maintain flight, the primary components of which are lift and drag. For steady flight, since we do not model a flapping wing, our goal is to calibrate energy by finding the coefficient of lift $C_L$ where the lift force balances that of gravity. We assume a mass of 0.08 kg, wing area of 0.0224 $m^2$ and average speed of 10 m/s for starlings \citep{Dial1997}\citep{Bruderer2001}.

\begin{equation}
\begin{split}
  |F_{lift}| = |F_{grav}| = \frac{1}{2} \rho v^2 A C_L & = m g \\
  C_L = 0.5714
\end{split}
\end{equation}

Using a lift-to-drag ratio of 3.3 for starlings from [Wither 1981] we can find $C_D = C_L / \frac{L}{D} = 0.1731$. In equilibrium the forward thrust force balances the total drag.  

\begin{equation}
\begin{split}
  |F_{thrust}| = |F_{drag}| = \frac{1}{2} \rho v^2 A C_D & = k_{power} \\
  k_{power} = 0.2373
\end{split}
\end{equation}

Applying these equilibrium parameters for $C_L$, $C_D$ and $k_{power}$ to our simulation, with social factors eliminated, results in average energy for induced power $P_{ind}$ of 7.8 W, combined parasitic and profile power of $P_{para}$ = 2.3 W and total mechanical power of $P_{mech}$ = 10.0 W. These values match the typical range of power observed for \textit{Sturnus vulgaris} \citep{Rayner1999} \citep{Ward2004}.

To calibrate Reynolds' simulation we must make assumptions since there are no aerodynamic forces in this model. Simulations typically converge to the minimum bird speed parameter which we set equal to our model at 10 m/s. Since there is no explicit orientation in Reynolds' model we assume the lift is vertical against gravity and drag is opposed to the forward velocity. These forces are emulated to compute and analyze energy; they do not alter the simulation itself.

We are interested in understanding the remaining energies used for forward acceleration and for turning. Our hypothesis is that Reynolds' model greatly exaggerates turning power since social forces applied in an arbitrary direction are not something a bird is realistically capable of producing. We define acceleration energy, $P_{accel}$, as any additional power applied after overcoming drag to increase or decrease the bird's speed, and turning energy, $P_{turn}$, as any additional power applied to turn toward a given target, as in Fig. \ref{fig_residual}. To analyze this, we compute these residual energies equivalently in both models by subtracting the lift, drag and gravity forces from the model-specific force that updates velocity ($F_{model}$=$F_{social}$ for Reynolds, $F_{model}$=$F_{flight}$ for ours). 

\begin{equation}
  F_{residual} = F_{model} - F_{lift} - F_{drag} - F_{grav} 
\end{equation}
\begin{equation}
  {\Delta}v = F_{residual} \: dt / mass 
\end{equation}
\begin{equation}
  P_{accel} = |F_{residual}| \: \left( {\Delta}v \cdot \hat{v} \right) 
\end{equation}
\begin{equation}
  P_{turn} = |F_{residual}| \: | {\Delta}v - \left( {\Delta}v \cdot \hat{v} \right) \hat{v} | 
\end{equation}

The component of residual force measured as work along the direction of motion gives $P_{accel}$, while the perpendicular component gives $P_{turn}$, as shown in Fig. \ref{fig_residual}. This framework enables the comparison of bird energy usage in both models equivalently, in real units of watts.

\begin{figure}[h]
  \begin{center}
  \includegraphics[width=0.65\textwidth]{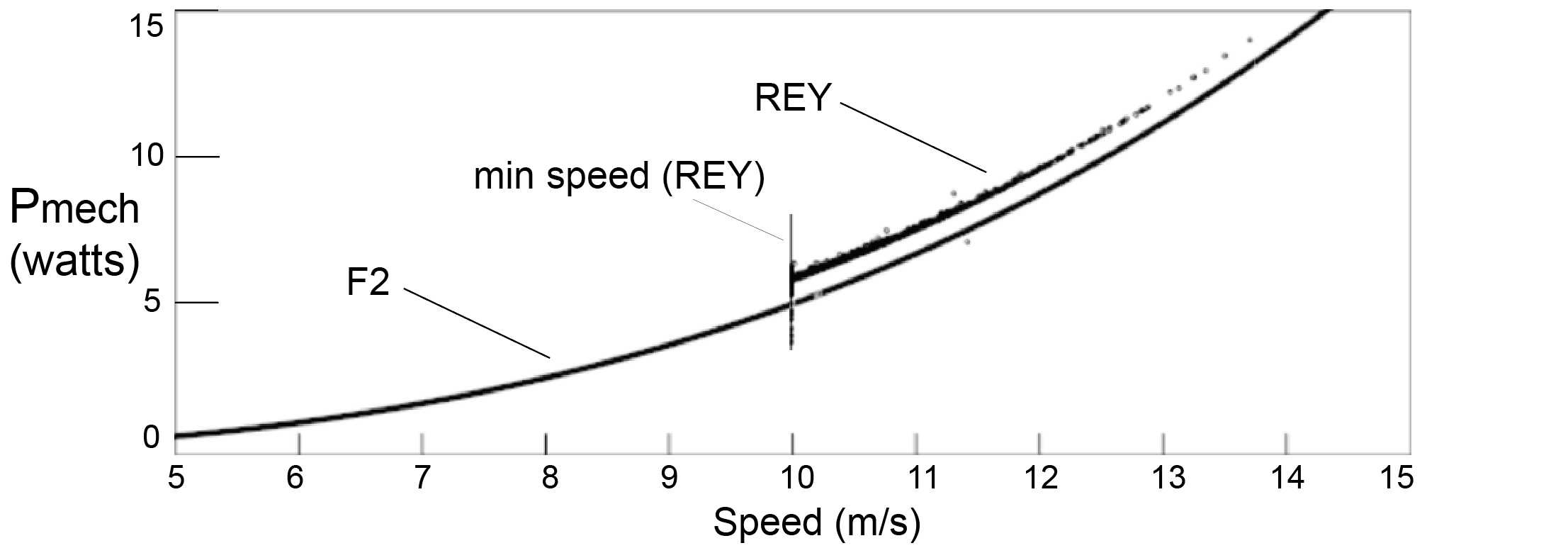}  
  \end{center}
  \caption{Total mechanical energy curves for both Flock2 (F2) and Reynolds (REY). Each point in this scatter plot for a simulation of 10,000 birds represents a bird at a specific velocity and mechanical energy output, $P_{mech}$. These curves exhibit the right-hand side of the theoretical U-shaped curve predicted for real birds \citep{Pennycuick1968}. In the Reynolds model the speed must be artificially limited to a minimum since bird velocities will converge to the minimum. In our model, F2, thrust determines the mean velocity. The F2 curve is composed of points yet appears solid in this plot as the power and velocities are more evenly and symmetrically distributed.}
  \label{fig_power}
\end{figure}

\begin{figure}[h]
  \begin{center}
  \includegraphics[width=0.55\textwidth]{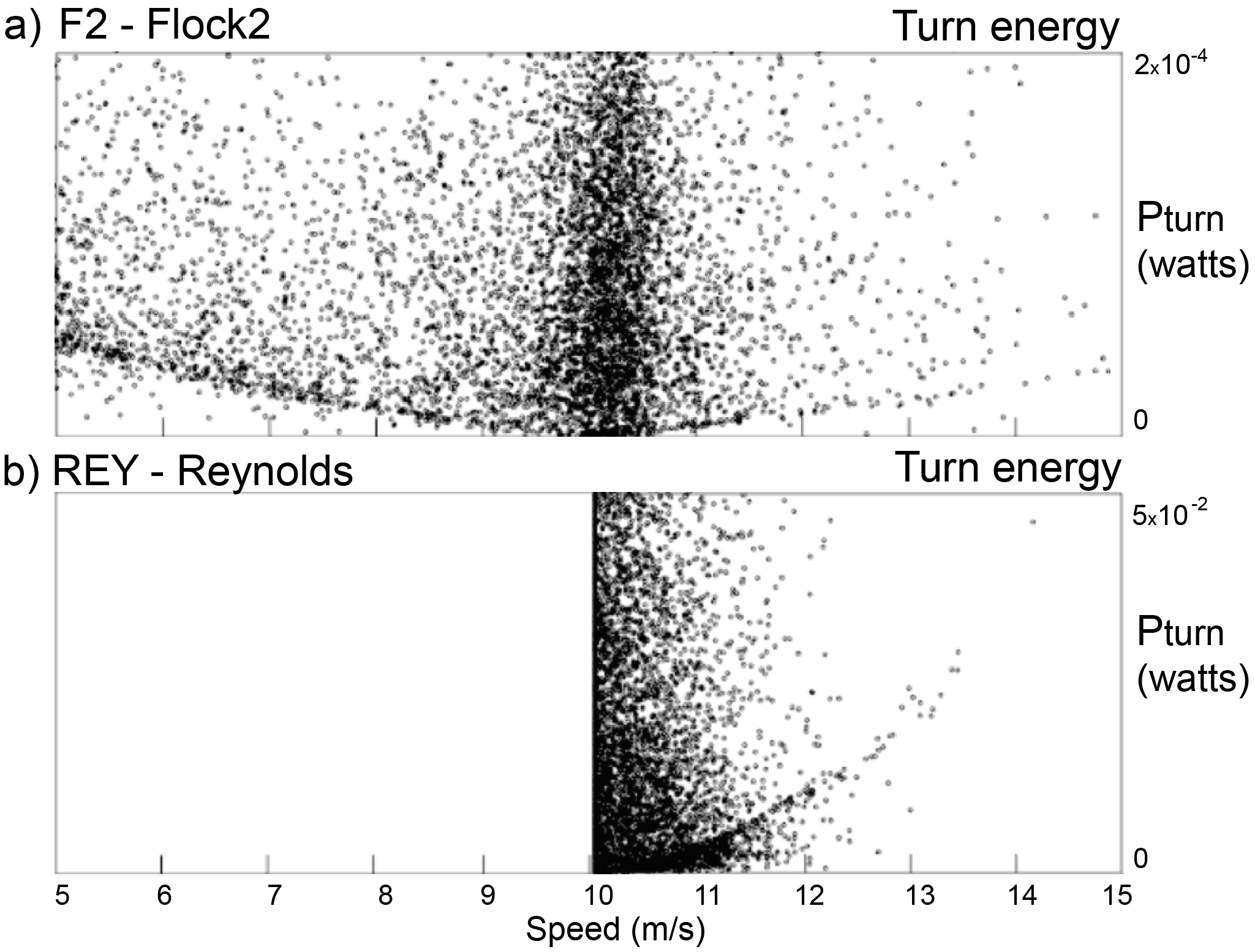}  
  \end{center}
  \caption{Total turning energy curves for the Flock2 model (top) and Reynolds model (bottom). Each point in this scatter plot for a simulation of 10,00 birds represents a bird at a specific velocity and $P_{turn}$. Notice that the F2 turn energies are two orders of magnitude smaller than the REY turn energies. Birds in the Reynolds model have a narrower range of speeds that converge to the minimum.}
  \label{fig_turns}
\end{figure}

Simulations were done with 10,000 birds in both models with the critical parameters $C_L$, $C_D$ and $k_{power}$ matching the equilibrium condition for a bird traveling 10 m/s. In Reynolds' model birds converge to the minimum speed parameter, therefore we artificially set this limit to 10 m/s for energy comparisons. In our Flock2 model, the minimum and maximum speeds are set to 5 m/s and 18 m/s respectively with the average bird velocity determined by $k_{power}$. Table 1 shows the bird energy output averaged over all birds. While the bird energies for lift, drag and acceleration are similar, the bird turning energies in Reynolds model were 39.5x (39,500\%) larger than our Flock2 model. Total mechanical energies and turning energies for all 10,000 birds are shown as scatter plots relative to bird velocity in Fig. \ref{fig_power} and \ref{fig_turns}, respectively. These energy results are compared to theories of bird flight in the discussion section.

\clearpage

\begin{figure*}[!htb]
  \includegraphics[width=\textwidth]{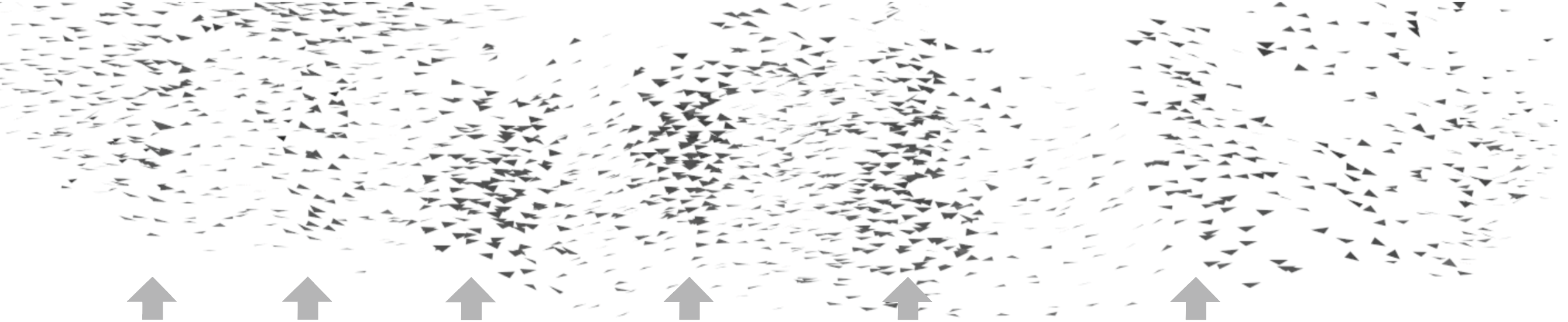}
  \caption{Detail of an orientation wave appearing in a Flock2 simulation of 10,000 birds. Birds are rendered as darts so that their aspect to the camera causes them to appear more or less opaque based on banking angle. Gray arrows highlight the crests of a moving wave where birds are presenting more wing area to the camera (appearing darker). The wave travels backward relative to bird motion at a speed much greater than the flock. See accompanying videos.}
  \label{fig_wave}
\end{figure*}

\begin{figure*}[!htb]
  \includegraphics[width=\textwidth]{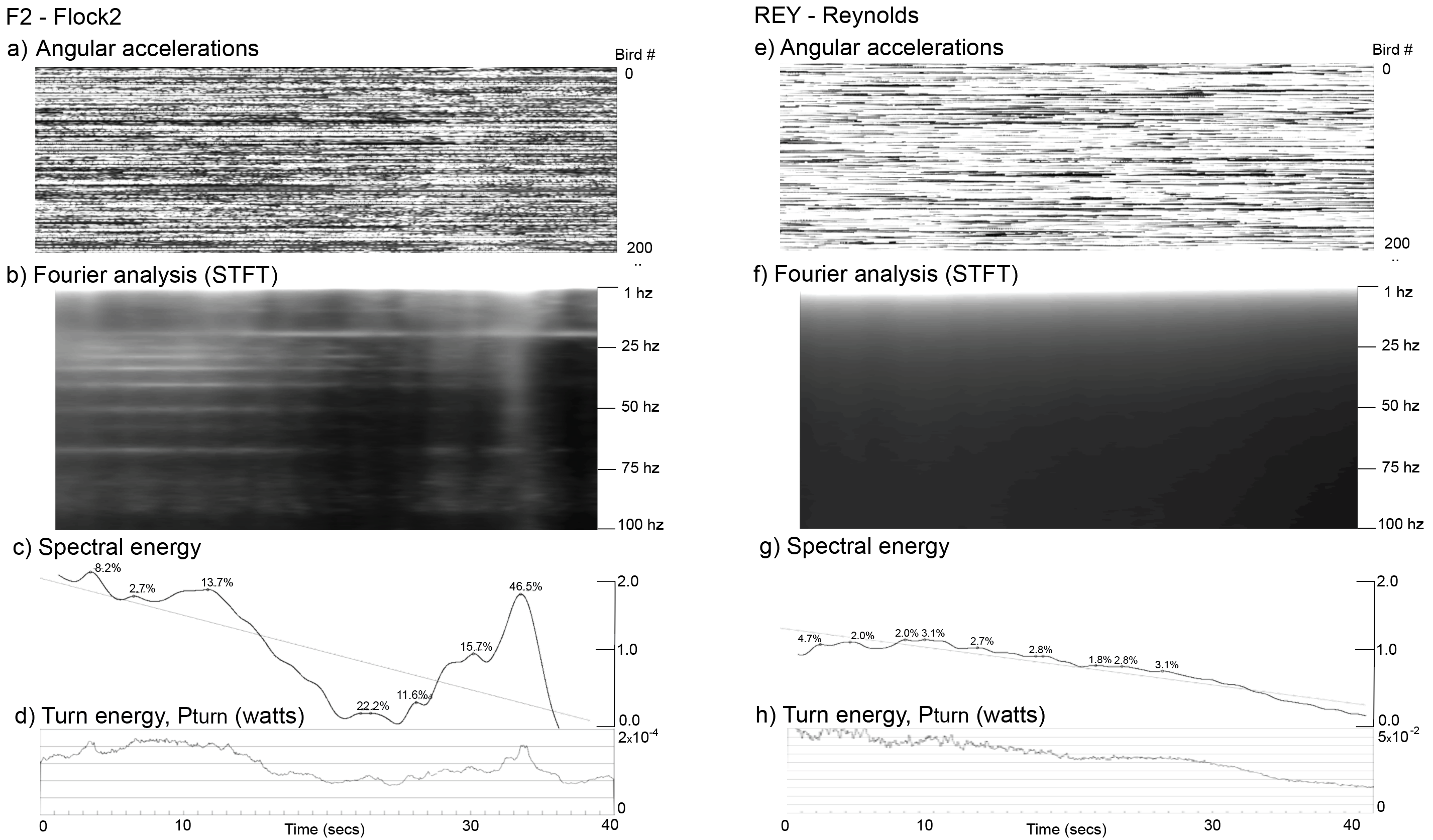}  
  \caption{Frequency analysis for the new Flock2 model (left) and Reynolds model (right). a) Angular accelerations for 200 birds are plotted here from a simulation of 10,000 birds and analyzed using b) Short-Time Fourier Transforms (STFT) to better understand the emergent low and high frequency oscillations observed in bird turns. c) The spectral energy of frequencies below 50 hz are accumulated to provide a picture of low frequency oscillations. Strong peaks were observed in F2 which directly corresponded to visible orientation waves in simulation videos. }
  \label{fig_freq}
\end{figure*}

\subsection{Frequency analysis}
\label{frequency}

Orientation waves are an emergent behavior in our simulation that occur as oscillations in bird orientation and which appeared in videos as rippling waves, shown in Fig. \ref{fig_wave}, that traverse from the point of impact of sub-flocks.

To examine this aspect we perform Fourier analysis on bird accelerations with the discrete-time Short-Time Fourier Transform (DT-STFT) to detect the frequency range of these oscillations [Sejdic 2009]. In this case, the magnitude of the angular acceleration of each bird, $|\alpha_i|$, is used as sample input. These inputs were scaled up by 100x in the Reynolds model prior to analysis to match the scale of Flock2 inputs, so that the STFT operates on data of equivalent scale across models. Analysis is conducted on both our Flock2 model and Reynolds model for a simulation of 10,000 birds.

Frequency analysis consists of the DT-STFT defined by a 1D real-to-complex Discrete Fourier Transform (DFT), implemented with the FFTW library, and applied over a moving window to isolate frequencies at each moment in time \citep{Jacobsen2003} \citep{Frigo2005}. This identifies the temporal frequency spectrum for a single bird at a moment in time. 

\begin{equation}
  X_i(f,t) = \sum_{n=-N/2}^{N/2} |\alpha_i(t+n)| \: H[n] e^{-j 2 \pi tk/N}
\end{equation}

The right side is a sliding window over the angular accelerations of a specific bird, $\alpha_i$, at time t with a Hanning window filter $H_{n}$ of width N, where N=512 is the number of frequency bins. The output X is a complex number for each frequency bin $f$, at each time $t$, for a given bird $i$. The magnitude squared of X gives the power of a particular frequency. 

\begin{equation}
  F_i(f,t) = |X_i(f,t)|^2
\end{equation}
  
The STFT detects the frequencies for a 1D signal; one bird. To summarize the entire flock we repeat the STFT for all birds (B) and accumulate the power at each frequency. If we had accumulated the input accelerations beforehand, the differing phases of the birds would cause signals to interfere and cancel out. By accumulating power after the STFT we can isolate and identify unique frequencies throughout the flock. If many birds are turning at 10 Hz, for example, the power of the 10 Hz frequency will be amplified regardless of where they are in the wave. 
  
\begin{equation}
  F_{total}(f,t) = \sum_{i=0}^{B} F_i(f,t)  
\end{equation}

The spectrograms of $F_{total}$ for Flock2 and Reynolds simulations of 40 seconds are shown in Fig. \ref{fig_freq}b and \ref{fig_freq}f, respectively. The simulation rate is 5 ms/frame therefore the sample rate is 200 Hz. The Nyquist frequency is 100 Hz and the bin size is 0.39 Hz/bin for N=512 bins. The spectrum exhibited by our Flock2 model has strong, specific, low and high frequency waves that are isolated in time.

Strong low frequency signals in Figure \ref{fig_freq}b were directly correlated with the observations of orientation waves in simulation videos, occurring most often when two sub-flocks interacted. This makes sense as the turning frequency must be low enough to be visually observable as a wave. To quantify this, we compute the total spectral energy over all frequencies below 50 Hz as shown in Fig. \ref{fig_freq}c and \ref{fig_freq}g.

\begin{equation}
  \text{SE}(t) = \sum_{f=0}^{50 \: hz} F_{total}(f,t)
\end{equation}
  
Peaks in total spectral energy with positive curvature and a deviation greater than 1\% from a fit line were detected and marked. These peaks were found to correspond very well with orientation waves observed in videos. We also confirm this result by noting that spectral power peaks, Fig. \ref{fig_freq}c and \ref{fig_freq}g, also correlate well with peaks in turning energy, $P_{turn}$, found in the earlier force analysis and plotted in Fig. \ref{fig_freq}d and \ref{fig_freq}h. This provides a cross-validation between the energy and frequency analyses that were performed.

\subsection{Sensitivity analysis}
\label{sensitivity}
Sensitivity analysis was conducted on various model parameters to understand how they impacted flock activity. These results are reported in Appendix B. Each simulation run was performed with 5000 birds, with a time step of 10 ms/frame, for 40 seconds giving a total of 4000 frames/run. A settling time of 5 seconds (500 frames) was allowed for the random initial positions of birds to settle into a nominal state. Parameters for number of birds, field of view,  boundary size and reaction rate were independently tested for spectral energy events (peaks >3\%), total low frequency power (0 < f <25 Hz) and total high frequency power (75 < f < 100 Hz).  

\begin{figure*}[!htb]
  \includegraphics[width=\textwidth]{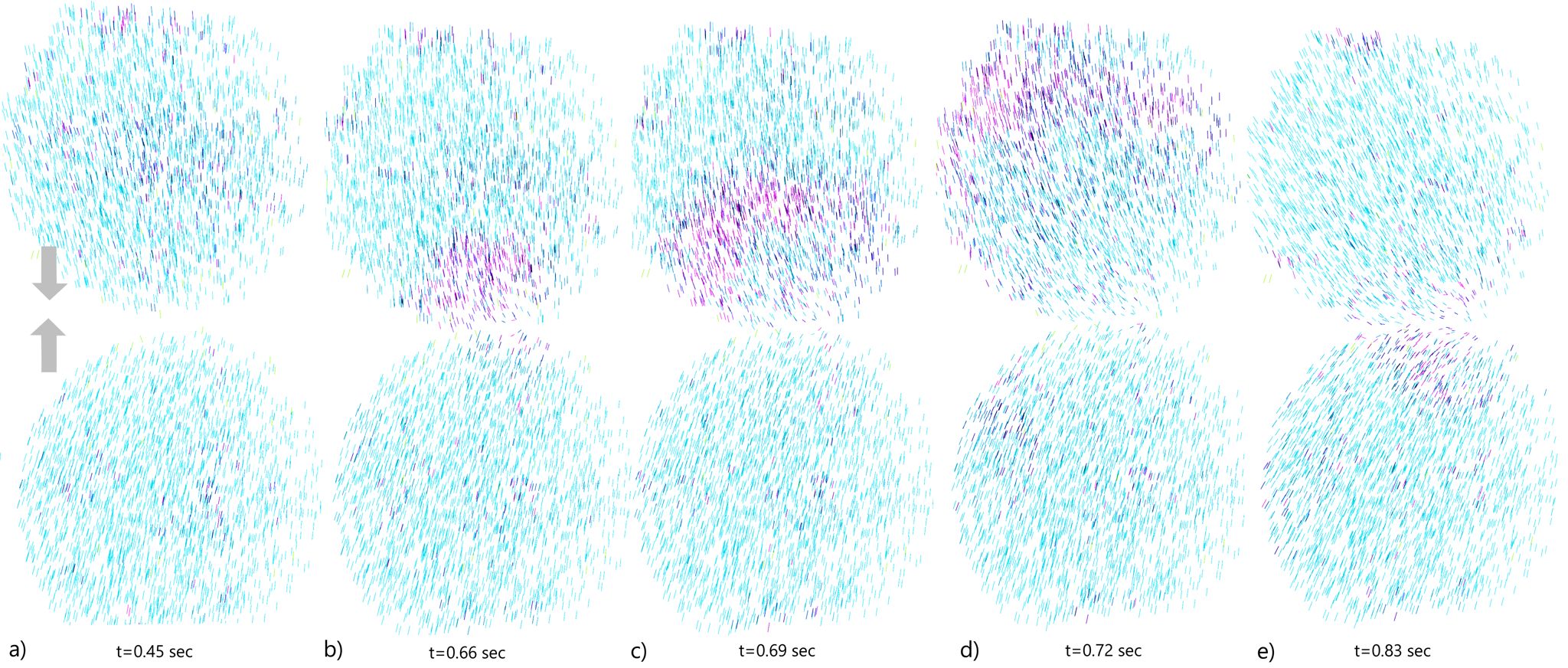}
  \caption{A collision experiment between two flocks of 5,000 birds with a timestep of 3 milliseconds. Purple represents birds with greater angular acceleration, i.e. rapidly turning. Gray arrows indicate initial velocities. a) Birds continue forward until b) the leading edge of birds responds to an upcoming collision by rapidly adjusting orientation, which c) propagates quickly to birds further from the edge, until d) reaching the entire flock. Angular acceleration waves may propagate e) asymmetrically at different times. On the order of one second the two flocks have established a new direction that avoids collision.}
  \label{fig_collide}
\end{figure*}

\subsection{Collision experiments}
\label{collisions}

Collision experiments were inspired by a live murmuration event seen in real starlings at 1:04 min in a public video by \citet{Clive2000}. At this moment the boundaries of two sub-flocks of starlings collide head on and then rebound without interpenetrating. We sought to replicate this. An experiment with two flocks of 5,000 birds was conducted with a timestep of 3 milliseconds over 1 second of simulation time in the Flock2 and Reynolds models. Upon approaching another flock in the F2 model, Fig. \ref{fig_collide}, leading birds turn quickly to avoid a collision which propagates rapidly through the entire flock. This high frequency response wave causes the two flock boundaries to avoid interpenetrating. Using the same time step for the Reynolds model, shown in the accompanying video, results in the two flocks merging and collapsing into a flattened disk rather than avoiding one another. In our model, the two flocks detect a collision, propagate turn information and resolve to a new direction without penetrating peripheral boundaries on the order of one second.

\begin{figure}
  \includegraphics[width=\textwidth]{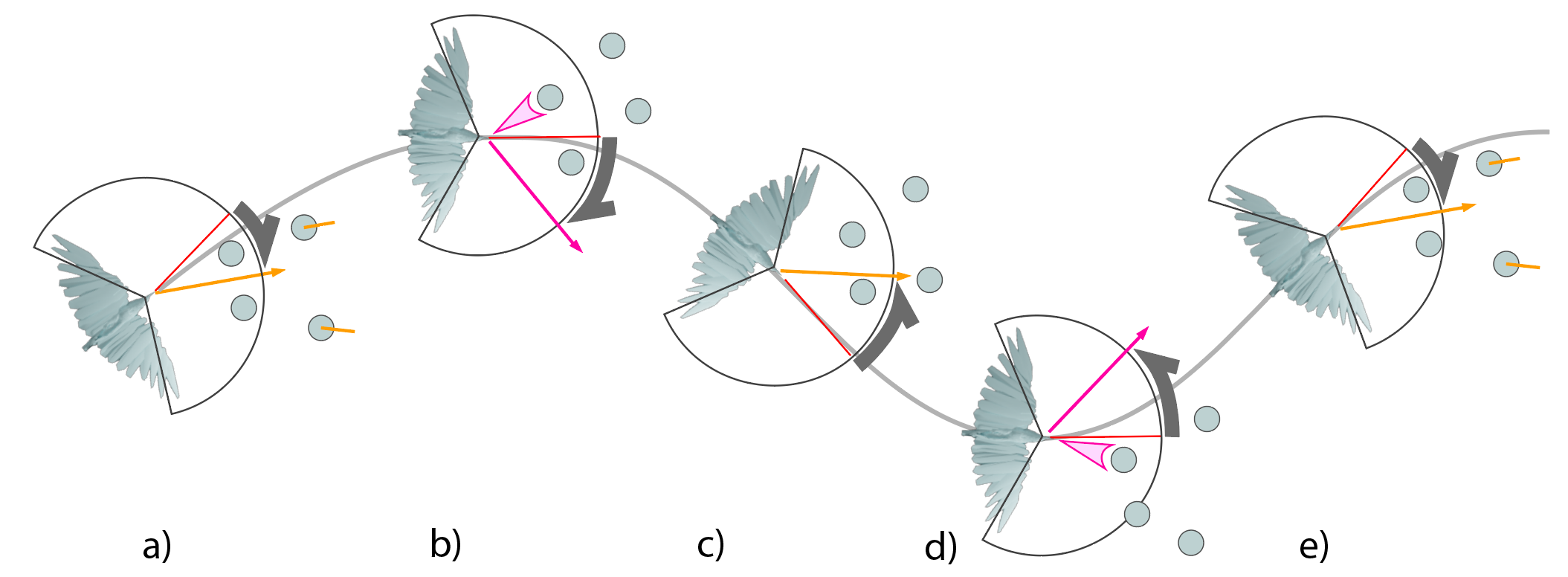}
  \caption{A plausible mechanism for emergent orientation waves. They arise in our model through the coupling of alignment (orange) and avoidance (purple) within the bird's field of vision, leading to an oscillatory turning as the bird approaches and turns away from escaping birds in front. See discussion.}
  \label{fig_escape}
\end{figure}
  
\clearpage 
\begin{figure*}[!htb]
  \includegraphics[width=\textwidth]{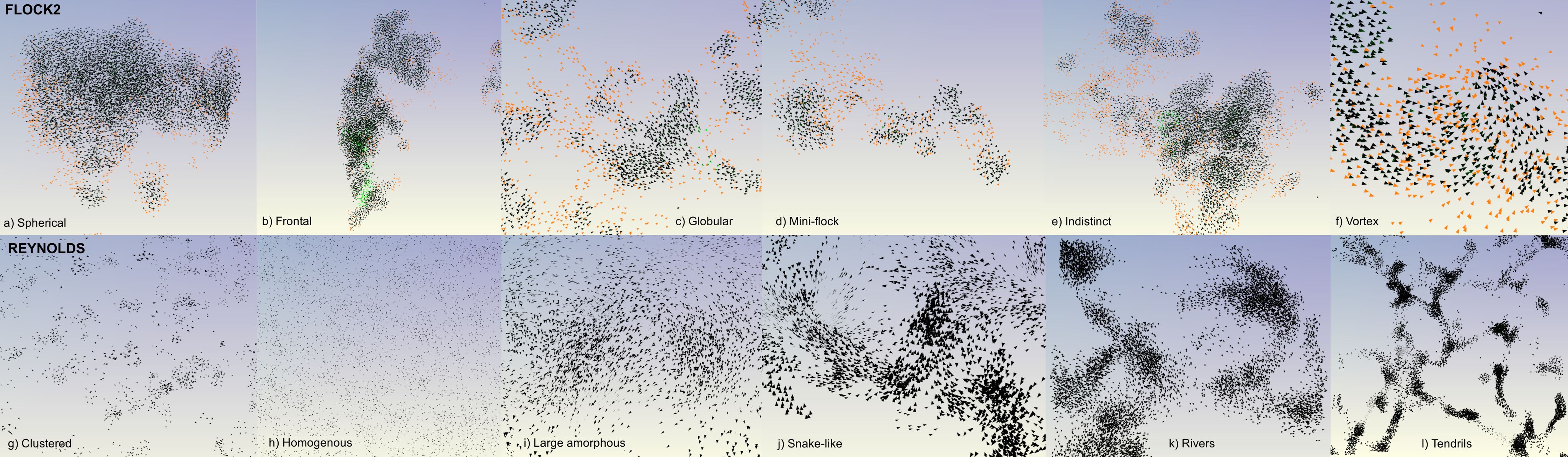}
  \caption{Flocking patterns exhibited by Flock2 (top) and Reynolds models (bottom). Ours exhibited a) spherical, b) frontal, c) globular, d) mini-flock, e) indistinct, and f) vortex patterns even within a single simulation run with 7 topological neighbors and the model parameters given in Appendix A. Reynolds' model, using the same 7 neighbor search radius, shows patterns described as g) clustered, h) homogenous, i) large amorphous, j) rivers, k) indistinct and l) tendrils, and has a more homogenous fluid-like appearance that converges to a single large flock moving uniformly, requiring multiple runs with different parameter settings.}
  \label{fig_patterns}
\end{figure*} 
 
\section{Discussion}
  
This work presents an improved simulation model, Flock2, for social flocking that exhibits novel behaviors for boundary cohesion, orientation waves and flocking patterns. We make this work available open source\footnote{Source code and supplemental video are available at: \href{https://github.com/ramakarl/Flock2}{\textcolor{blue}{https://github.com/ramakarl/Flock2}}} so that the community may improve on and benefit from these computational experiments. To our knowledge this is the first model to employ social orientation where the behaviors of avoidance, alignment and cohesion are implemented as turn targets rather than force vectors. Aerodynamic forces are computed as 3D vectors for lift, drag, thrust and gravity, yet oriented solely through actuation of the target heading.
  
Emergent orientation waves were observed for the first time in these simulations without explicit coding or external triggers. We distinguish between orientation waves, without predators, and agitation waves which require predators as a trigger, and suggest that general orientation waves may appear in real starling murmurations in the absence of explicit predators. Our work does not refute the fact that agitation waves are the result of an escape response to predator attacks \citep{Procaccini2011}, yet it also validates that aerial displays of orientation waves may not require an explicit attack by a predator \citep{Carere2009}. Rather, orientation waves might also occur more generally when the flock self-intersects or when the flock density suddenly triggers oscillating motions within the flock itself.

Our method does not require a transmission mechanism \citep{Hemelrijk2015} for orientation waves. The mechanism of action may be explained as a feedback coupling of the competing turning forces of alignment and avoidance within the bird’s field of vision (Fig. \ref{fig_escape}). When only force-based avoidance is used, as in previous models, spontaneous orientation waves do not occur. In the Flock2 model, a bird traveling with a flock (Fig. \ref{fig_escape}a) may approach too rapidly if birds ahead have reoriented, causing it to turn in avoidance (Fig. \ref{fig_escape}b). Once a safer orientation is found, the bird is compelled again to return to the flock (Fig. \ref{fig_escape}c), only needing to dodge again (Fig. \ref{fig_escape}d). Sensitivity analysis shows this is coupled to the bird's field of view (Figure B.13).

Flocking patterns in our model are complex and appear to better mimic those in nature, Figure \ref{fig_patterns}. \citet{Carere2009} classifies the aerial flocking patterns of European starlings into the categories of indefinite, globular, miniflock, frontal, ovoid, snake-like, singleton, spherical, linear, V-shaped, and planar. Many of these patterns could be observed in our model even within a single simulation run with fixed model parameters. \citet{Carere2009} also found that the indeterminate, ovoidal and miniflock patterns were the most common shapes over 1537 observations, patterns which also occurred most frequently in our simulations. 
  
Our model exhibits large, dense globular flocks even with the perceptual range limited to seven topological neighbors in all experiments. Flock2 generates flocks with thousands of birds that split into several mini-flocks and re-merge continuously (see supplemental video). With a similar constraint, Reynolds' model results in non-globular flocks that are more fluid-like with patterns that are clustered, flowing or tendril-like \citep{Mototake2015}. While Reynolds' model frequently exhibits spherical mini-flocks of 10-20 birds, we were only able to generate large globular flocks with more than 200 birds in that model when using an indefinite (or very large) search radius.

Introduction of the peripheral boundary term relies on the fact that flocking is an evolved response \citep{Wood2007} \citep{Beauchamp2004} \citep{Kwasnicka2011}. Bird flocks form a cohesive periphery in our model not because a predator is immediately present but because we assume they have evolved in the predator-prey relationship to aggregate for protection. Proximity to the periphery, or exposure, is the only variable required to model this. This explanation is more satisfying than one where attraction to a roost is used to provide in-flight flock cohesion. Without some peripheral bounding term, Flock2 converges to long echelon shaped flocks while Reynolds' model converges to a single large, homogeneous flock moving uniformly as confirmed by others \citep{Bajec2007}. 
  
\section{Conclusions \& Future work}  

By decoupling social and aerodynamic factors, and requiring that social orientation targets mimic the vision system to provide only heading goals for an underlying aerodynamic model for the mechanics of flight, we believe this system is more realistic with respect to both real bird flight and aerial flocking. Nonetheless an important limitation is that we do not yet model flapping wings which should require only a change to the aerodynamic model.

A more complete model would consider both the internal and external power in flight including both mechanical and metabolic work. This requires a more complete functional model of bird metabolism and would be an interesting direction for future research. 

Orientation waves are examined with energy and frequency analysis, yet a deeper understanding of their arisal would be beneficial in both computational models and real birds. An important question raised by this theoretical work is the prevalence of orientation waves in large murmurations of real birds in the absence of predators, for which there is still little experimental research since the focus of field monitoring has been on predator-prey dynamics. This may require high resolution, high speed cameras to provide a good control in ensuring there are no predators actively flying in a very large flock. 

In the future we hope to implement predators in order to compare different triggers, types of waves, and mechanisms of propagation. Since predators are known to trigger agitation waves we would expect them to emergently trigger orientation waves.

We provide the Flock2 model as open source software in the hope that it contributes to observational and computational biology by enabling further research in aerial flocking, bird flight, control theory, and swarm dynamics.

\section{Funding}
This research did not receive any specific grant from funding agencies in the public, commercial, or not-for-profit sectors.


\appendix
\label{apxA}

\clearpage
  
\begin{table}[ht]
  \section{Model Parameters}
  \label{tab:results}
  \begin{tabular}{lrrr}
  \toprule
  Parameter & Symbol & Value & Units  \\
\midrule
\textit{Social parameters} & & \\
Avoidance strength & $k_{avoid}$ & 0.02 & \\
Alignment strength & $k_{align}$ & 0.60 & \\
Cohesion strength & $k_{cohes}$ & 0.004 & \\
Boundary strength & $k_{bound}$ & 0.10 & \\
Boundary size & $B$ & 20 & \# birds \\
Neighbor influence & $N_{topo}$ & 7 & \# birds \\
Field of view & $fov$ & 290 & degrees \\
\textit{Flight parameters} & & \\
Mass & $m$ & 0.08 & kg \\
Wing area & $A$ & 0.0224 & $m^2$ \\
Coeff. of lift & $C_L$ & 0.5714 & \\
Coeff. of drag & $C_D$ & 0.1731 & \\
Lift/drag ratio & L/D & 3.3 & \\
Reaction speed & $k_r$ & 250 & msec \\
Dynamic stability & $k_s$ & 70\% & \\
Initial velocity & $v_{init}$ & 10 & $m/s$ \\
Min velocity & $v_{min}$ & 5 & $m/s$ \\
Max velocity & $v_{max}$ & 18 & $m/s$ \\
Thrust & $k_{power}$ & 0.2373 & N \\
\textit{Environmental parameters} & & \\
Air density & $\rho$ & 1.225 & $kg/m^3$ \\
Gravity & g & 9.8 & $m/s^2$ \\
\textit{Simulation \& testing} & & \\
Time step & $dt$ & 5 & msec/frame \\
Settle time & $t_{start}$ & 5 & sec \\
Run time & $t_{sim}$ & 20 & sec \\
FFT size, section \ref{frequency} & N & 512 & bins\\
Sample rate & s & 200 & samples/sec \\
Nyquist freq. & $N_f$ & 100 & hz \\
Bin size & $B_f$ & 0.39 & hz/bin \\
\textit{Reynolds' parameters} & & \\
Avoidance strength & $R_{avoid}$ & 0.16 & \\
Alignment strength & $R_{align}$ & 0.60 & \\
Cohesion strength & $R_{cohes}$ & 0.08 & \\
  \bottomrule
  \end{tabular}
\end{table}

\clearpage
\begin{center}
\begin{figure*}[h!]
  \section{Sensitivity Analysis}
  \includegraphics[width=0.85\textwidth]{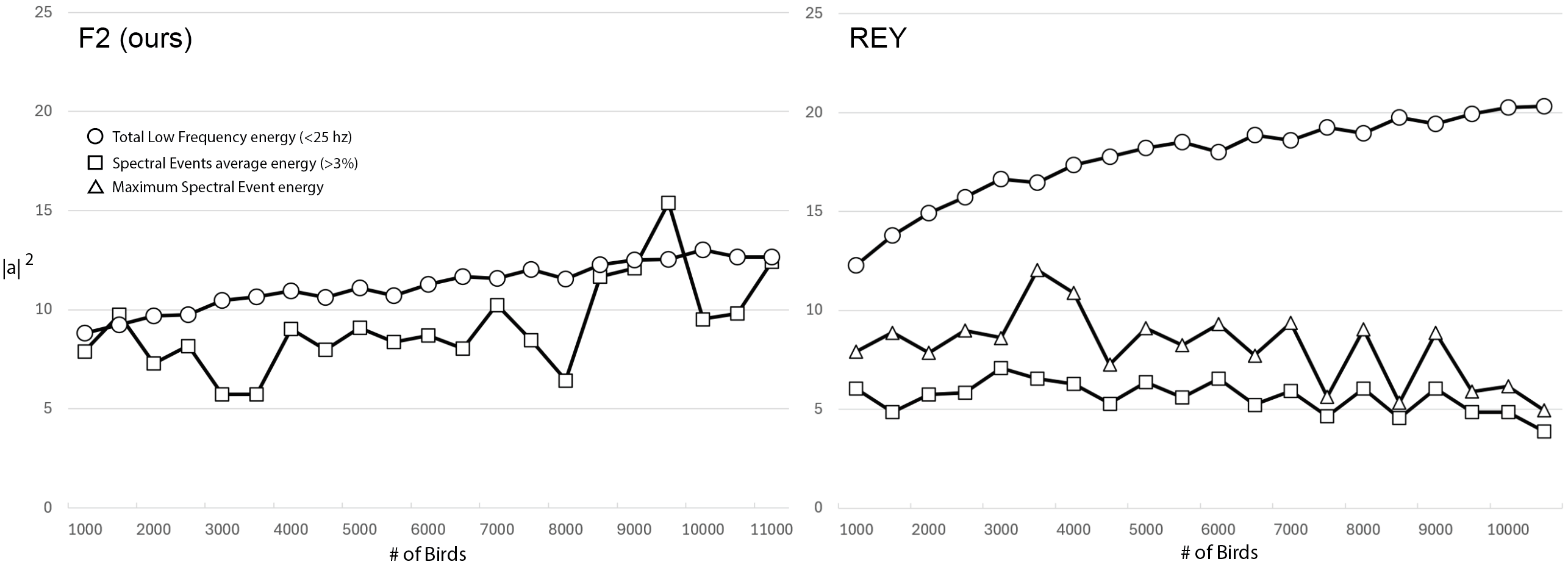}  
  \caption{Number of birds versus spectral events and energy for F2 and REY. The spectral events, corresponding to orientation waves, are peaks in SE over 3\% with average energy shown (squares). The total low frequency energy (<25 hz, circle) is also shown. The y-axis is in units of magnitude of angular acceleration squared. }
\end{figure*}
\begin{figure*}[h!]
  \includegraphics[width=0.85\textwidth]{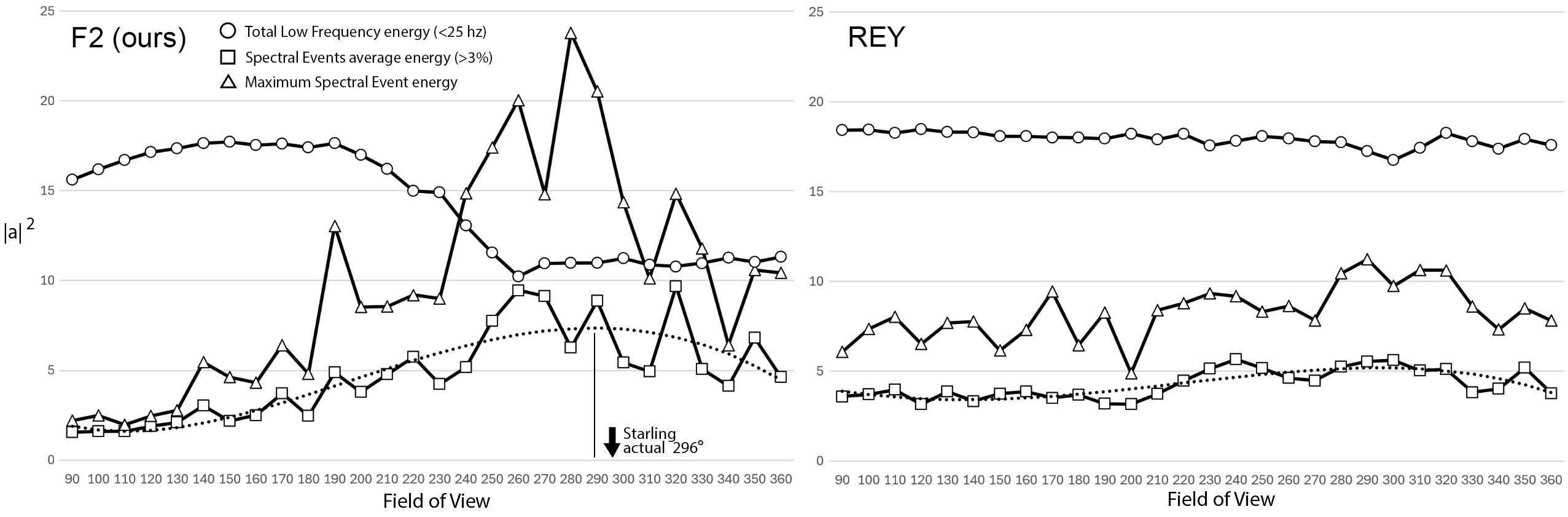}  
  \caption{Field of view versus spectral events and energy for F2 and REY. The spectral events, corresponding to orientation waves, are peaks in SE over 3\% with average energy shown (squares). The total low frequency energy (<25 hz, cirlces) is also shown. The y-axis is in units of magnitude of angular acceleration squared. The strongest event energy corresponds very well to the actual field of view of \textit{Sturnus vulgaris}, 296 degrees.}
\end{figure*}
\begin{figure*}[h!]
  \includegraphics[width=0.85\textwidth]{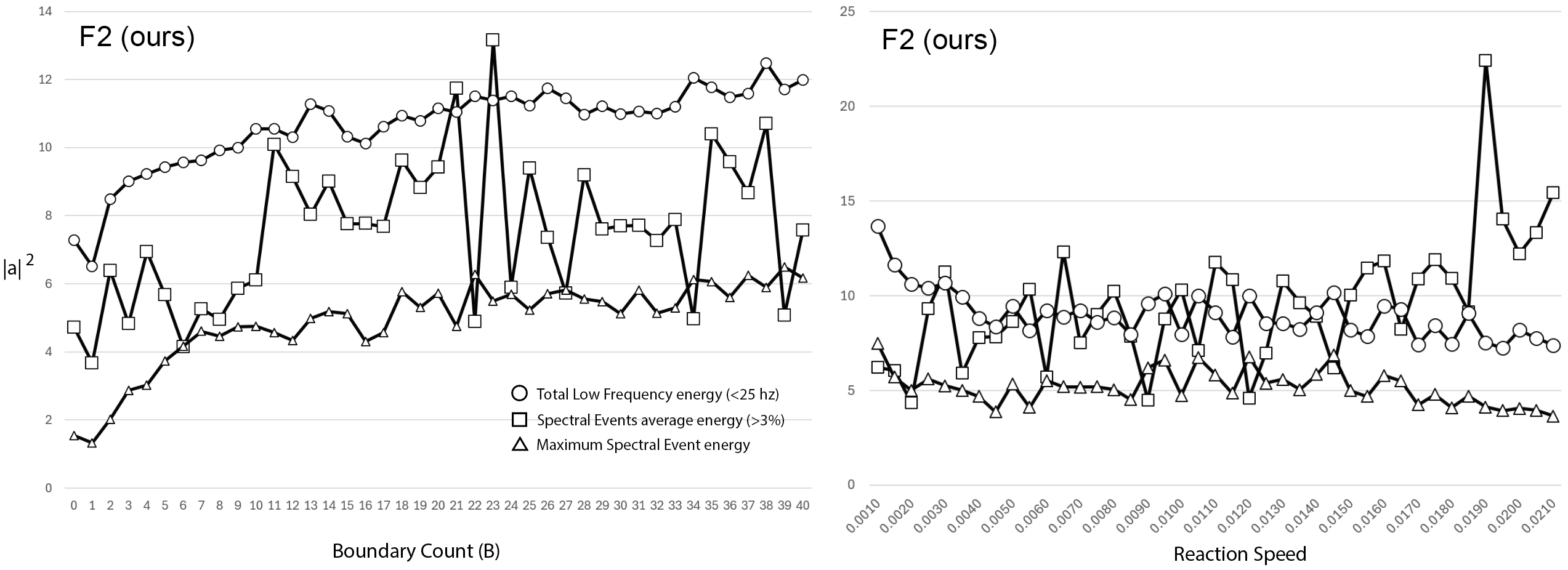}  
  \caption{Boundary count term (B), left, and reaction speed, right, plotted against spectral events and energy in F2. Total turning energies in low frequency (<25 hz, cirlces) and high frequency (>75 hz, triangles) are shown. As the peripheral boundary term increases (width of boundary in number of birds B) the total event energy, low frequency, and high frequency energy all increase.}
\end{figure*}
\end{center}

\clearpage

\begin{figure*}[h!]
  \section{Visual results}
  \includegraphics[width=0.85\textwidth]{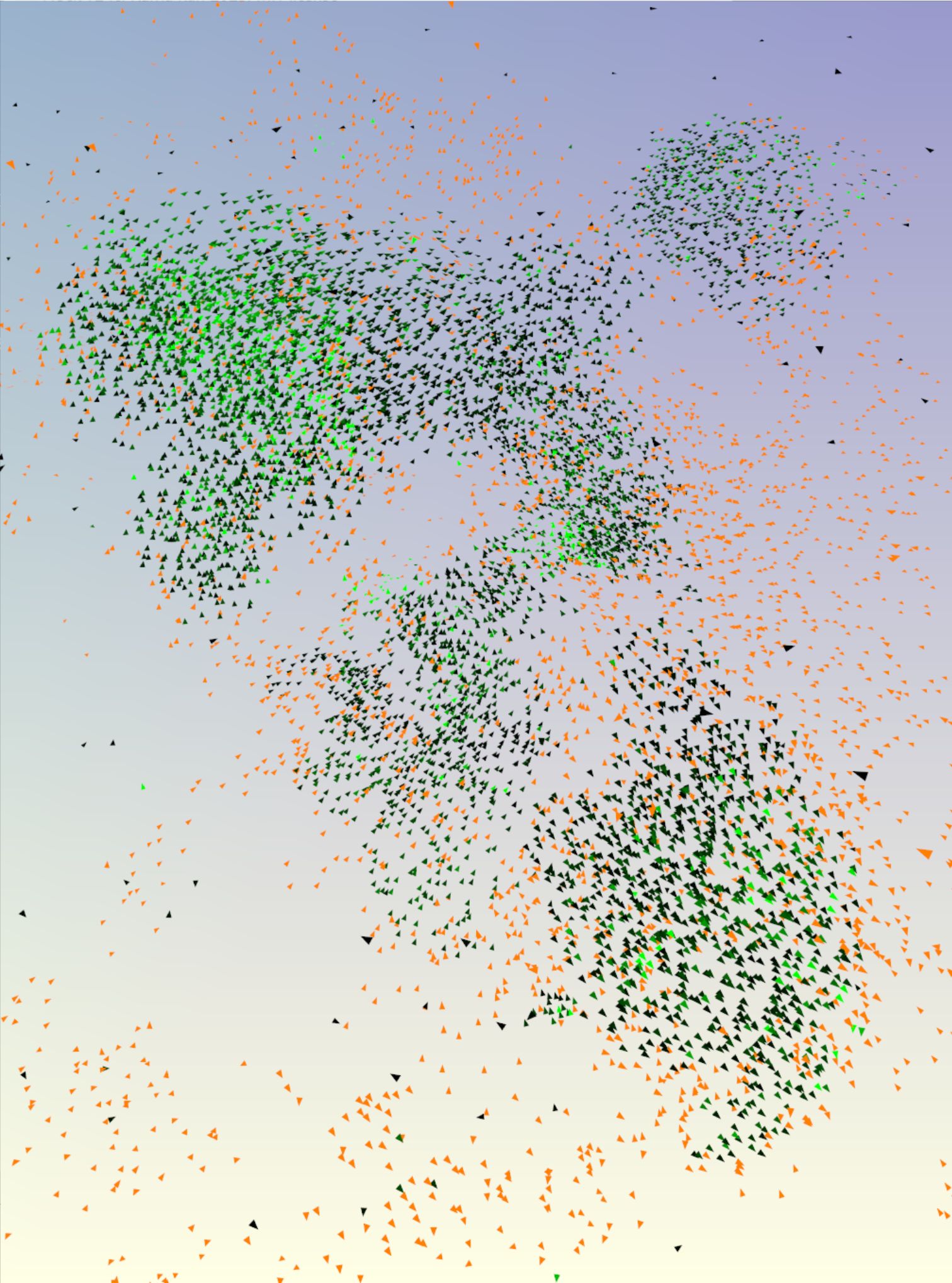}  
  \caption{Visual output of Flock2. Birds are rendered as triangles with orientation as required to exhibit orientation waves based on area of the wings presented to the camera. Birds influenced by the peripheral boundary term are colored orange. Extreme changes in angular acceleration are visualized in green.}
\end{figure*}



\clearpage

\bibliographystyle{unsrtnat}
\bibliography{ms}  



\end{document}